\documentclass{gGAF2e}
\usepackage{amsmath,amssymb}
\usepackage{graphicx}
\usepackage{amssymb}
\usepackage{amsfonts}
\usepackage[normalem]{ulem}
\usepackage{color}
\usepackage[makeroom]{cancel}

\newcommand{\per}{\bm{\nabla}_{\perp}}
\newcommand{\nabper}{\nabla_{\perp}}
\newcommand{\para}{\bm{\nabla}_{\parallel}}

\newcommand{\dvper}{\delta u_{\perp}}
\newcommand{\dvpara}{\delta u_{\parallel}}

\newcommand{\Bper}{\delta B_{\perp}}
\newcommand{\Bpara}{\delta B_{\parallel}}

\newcommand{\ep}{\varepsilon_1}

\newcommand{\ept}{\varepsilon_{2}}

\newcommand{\bfzhat}{\mbox{\boldmath $ {\hat z}$}}
\newcommand{\bfxhat}{\mbox{\boldmath $ {\hat x}$}}

\newcommand{\bfB}{\mbox{\boldmath$B$}}

\newcommand{\bff}{\mbox{\boldmath$f$}}

\newcommand{\bfb}{\mbox{\boldmath$b$}}

\newcommand{\bfu}{\mbox{\boldmath$u$}}

\newcommand{\bfF}{\mbox{\boldmath$F$}}

\newcommand{\drm}{\mathrm{d}}

\newcommand{\irm}{\mathrm{i}}
\newcommand{\Grm}{\mathrm{G}}
\newcommand{\Orm}{\mathrm{O}}

\newcommand{\dz}{{\mathrm{d}}z}

\makeatletter
\newsavebox\myboxA
\newsavebox\myboxB
\newlength\mylenA

\newcommand*\xoverline[2][0.75]{%
    \sbox{\myboxA}{$\m@th#2$}%
    \setbox\myboxB\null
    \ht\myboxB=\ht\myboxA%
    \dp\myboxB=\dp\myboxA%
    \wd\myboxB=#1\wd\myboxA
    \sbox\myboxB{$\m@th\overline{\copy\myboxB}$}
    \setlength\mylenA{\the\wd\myboxA}
    \addtolength\mylenA{-\the\wd\myboxB}%
    \ifdim\wd\myboxB<\wd\myboxA%
       \rlap{\hskip 0.5\mylenA\usebox\myboxB}{\usebox\myboxA}%
    \else
        \hskip -0.5\mylenA\rlap{\usebox\myboxA}{\hskip 0.5\mylenA\usebox\myboxB}%
    \fi}
\makeatother

\newcommand{\bmu}{\bm{u}}
\newcommand{\bmB}{\bm{B}}
\newcommand{\bmnabla}{\bm{\nabla}}
\newcommand{\db}{\bmB_0}

\newcommand{\hp}{\hat{p}}

\newcommand{\hr}{\hat{\rho}}

\begin{document}
\title{{\textit{Incorporating Velocity Shear into the Magneto-Boussinesq Approximation}}}

\author{JORDAN A. BOWKER, DAVID W. HUGHES{$^\ast$}\thanks{$^\ast$Corresponding author. Email: d.w.hughes@leeds.ac.uk} and EVY KERSAL\'E \\
\vspace{6pt} Department of Applied Mathematics, University of Leeds, Leeds LS2 9JT, UK\\ }
\received{Received Date}
\maketitle

\begin{abstract}
Motivated by consideration of the solar tachocline, we derive, via an asymptotic procedure, a new set of equations incorporating velocity shear and magnetic buoyancy into the Boussinesq approximation. We demonstrate, by increasing the magnetic field scale height, how these equations are linked to the magneto-Boussinesq equations of Spiegel and Weiss (1982).


\end{abstract}

\markboth{Velocity shear in the Boussinesq approximation}{J.A.~Bowker et al.}

\section{Introduction\label{sec:intro}}

Instabilities driven by magnetic buoyancy have been studied over a number of years, with particular emphasis given to their role in disrupting a strong, predominantly toroidal magnetic field in the solar interior \citep[see, for example, the review by][]{Hughes_2007}. For a variety of (essentially unrelated) reasons, it has been suggested that the bulk of the Sun's magnetic field is stored at the base of, or just below, the convection zone. From estimates of the rise times of magnetic flux tubes through the convection zone, \citet{Parker_1975} argued that it would be difficult to confine the magnetic field for times comparable with the solar cycle period unless the dynamo operated only in the `very lowest levels of the convective zone'. \citet{Golub_etal_1981}  \citep[see also][]{SW_1980} proposed a similarly deep-seated layer of toroidal field, but from arguments based instead on the expulsion of magnetic fields by convective motions. Perhaps the most compelling evidence for pinning down the location of the solar toroidal field comes from the discovery, by helioseismology, of the solar tachocline, a thin region of strong radial and latitudinal velocity shear, sandwiched between the convective and radiative zones \citep{Schou_etal_1998}. Although there is little consensus on exactly how the solar dynamo operates, it is generally agreed that toroidal field is wound up from a relatively weak poloidal ingredient via strong differential rotation (the $\omega$-effect of mean field dynamo theory). Consequently, the tachocline becomes the natural location for a deep-seated, predominantly toroidal magnetic field.

Given this, it is natural to seek to build upon previous studies of magnetic buoyancy instabilities by incorporating the effects of a velocity shear. Using the energy principle, \citet{TH_2004}, extending the results of \citet{Adam_1978}, obtained necessary conditions for the ideal (diffusionless) linear instability of a magnetohydrodynamic (MHD) state with aligned horizontal flow and magnetic field, each stratified arbitrarily in the vertical direction. From a different perspective, \citet{VB_2008} considered the fully nonlinear evolution of magnetic buoyancy instabilities in a magnetic layer generated through the stretching of an initially vertical magnetic field by a horizontal, depth-dependent shear flow.

Instability due to magnetic buoyancy is an inherently compressible phenomenon, with the magnetic pressure playing the crucial role in reducing the local density of the gas. Thus, most studies of the instability have employed the equations of fully compressible MHD. However, just as convection of a compressible fluid can, under certain circumstances, be treated within the almost-incompressible Boussinesq approximation \citep{SV_1960}, so can magnetic buoyancy be incorporated into a similar magneto-Boussinesq approximation (Spiegel and Weiss 1982; hereinafter SW82). Such approximations afford a simplification of the governing equations and thus aid both theoretical and numerical analysis. Our aim in this paper is to incorporate the effects of a velocity shear into the magneto-Boussinesq equations, self-consistently and in such a way that the influence of the shear is comparable with that of the magnetic buoyancy instability.

The equations of the Boussinesq approximation for a compressible fluid were derived in the classic paper of \citet{SV_1960}, who considered thermal convection of a layer of fluid subject to two important assumptions: the first is that the depth of the fluid layer is much smaller than the scale height of any thermodynamic quantity; the second is that motion-induced fluctuations in density, temperature and pressure do not exceed their static variation. The first assumption is a statement about the basic state, the second is an eminently reasonable supposition that can be verified \textit{a posteriori}. Under these assumptions, the governing equations simplify considerably. In particular, the fluid is treated as incompressible, with density variations neglected except in the buoyancy term in the equation of motion; furthermore, fluctuations in the pressure are small --- a reflection of the low Mach number --- and thus density variations are directly proportional to variations in temperature.

For problems such as magnetoconvection, magnetic fields can be incorporated into the Boussinesq approximation in a straightforward manner \citep[see, for example,][]{PW_1982}. The field enters through the induction equation and via the Lorentz force in the momentum equation; variations in magnetic pressure are assumed to have no influence on density fluctuations. Including the effects of magnetic buoyancy is however a more subtle procedure. SW82 considered the problem of the instability of a stratified, horizontal magnetic field with scale height $H_B$ very large compared with the layer depth $d$.  The crucial ordering is now one in which variations in the \textit{total} pressure (gas $+$ magnetic) are small; this has implications for all the governing equations. In the momentum equation, density fluctuations are related to variations in both the temperature and the magnetic pressure; similarly, variations in magnetic pressure enter into the energy equation. The velocity is, to leading order, incompressible. However, it becomes necessary to include the next order correction to $\bmnabla {\bm \cdot} \bmu$ in the induction equation; in standard notation this then takes the form
\begin{equation}
\frac{\partial \bmB}{\partial t} + \bmu {\bm \cdot} \bmnabla \bmB = \bmB {\bm \cdot} \bmnabla \bmu - \frac{w}{H_\rho} \bmB + \eta \nabla^2 \bmB,
\end{equation}
where $H_\rho$ is the density scale height of the basic state. The final, and extremely important feature to note is that within the magneto-Boussinesq approximation, magnetic buoyancy is relevant only for modes of a certain horizontal scale. In particular, when considering the stability of an equilibrium state with a unidirectional horizontal field, magnetic buoyancy is of significance for perturbations with a long ($\Orm (H_B)$) length scale in the direction of the imposed field. One consequence of this is that the magnetic field is not exactly solenoidal; $\bmnabla {\bm \cdot} \bmB = 0$ only to $\Orm (d/H_B)$, an approximation that is however consistent with the overall level of approximation introduced in the magneto-Boussinesq approximation.

In an approach complementary to that of SW82, Corfield (1984) (hereinafter C84) re-derived the magneto-Boussinesq equations through a formal scaling analysis, expanding all variables in terms of the two small parameters of the system: $d/H$, where $H$ denotes any of the scale heights (all comparable), and $\delta \rho/\rho_0$, the ratio of fluctuations in density to a representative value.

Our aim in this paper is to incorporate the effects of a shear flow into the magneto-Boussinesq approximation. As explained in Appendix~\ref{ap:noshear}, if this is done in what might be considered the obvious fashion --- namely with the shear flow of the same order as the velocity perturbations in Corfield's ordering --- then the influence of the shear has no bearing on the onset of instability. Thus, in order to consider a regime in which a shear flow may interact with the magnetic buoyancy instability, it becomes necessary to consider in some detail the magnitudes of the imposed magnetic field and the velocity shear flow, together with their gradients, as well as the horizontal scale of the perturbations.

In section~\ref{sec2} we present a derivation of the scalings inherent to the magneto-Boussinesq approximation in the absence of an imposed shear flow; the derivation is along similar lines to that of \citet{Corfield_1984}, though we are more explicit in stating the underlying physical assumptions. In section~\ref{sec3} we explore the orderings of the imposed shear flow and magnetic field that are necessary in order to accommodate the effects of magnetic buoyancy and velocity shear on the same footing. Following this, section~\ref{sec4_1} contains the main result of the paper, the derivation of asymptotically consistent magneto-Boussinesq equations incorporating velocity shear; the crucial differences with the equations of C84 are discussed in section~\ref{sec4_2}. In section~\ref{sec5} we explore these differences systematically by explaining how the various scalings change with the magnitude of the magnetic field scale height, thus providing a transition between the equations of \citet{Corfield_1984} and our new set of equations. The concluding discussion is contained in section~\ref{sec6}.

\section{The magneto-Boussinesq approximation \label{sec2}}
In standard notation, the magnetohydrodynamic equations for a perfect gas are
\begin{subequations}
\label{eq:MHD}
\begin{align}
\label{eq:basic1}\upartial_t \rho + \bmnabla {\bm \cdot}(\rho \bmu) &= 0, \\
\bmnabla {\bm \cdot} \bmB &= 0, \\
\rho\left(\upartial_t + \bmu {\bm \cdot} \bmnabla \right) \bmu = -\bmnabla \varPi &- g \rho \bfzhat + \mu_0^{-1} \bmB {\bm \cdot} \bmnabla \bmB + \bm{F} + \bmnabla {\bm \cdot} \bm{\tau},\\
\left(\upartial_t + \bmu {\bm \cdot} \bmnabla \right)\bmB = \bmB {\bm \cdot} \bmnabla \bmu &- \bmB (\bmnabla {\bm \cdot} \bmu) + \eta \nabla^2\bmB,\\
\rho c_p\left(\upartial_t + \bmu {\bm \cdot} \bm{\nabla} \right) T - \left(\upartial_t + \bmu {\bm \cdot} \bm{\nabla} \right)p &= K \nabla^2 T + \eta \mu_0^{-1} (\bm{\nabla} \bm{\times} \bmB)^2 + \Phi ,\\
p &= R \rho T , \label{eq:MHD_f}
\end{align}
\end{subequations}
where $\bfzhat$ is the unit vector in the vertical direction, $\varPi$ is the total pressure, consisting of the sum of the gas pressure $p$ and the magnetic pressure $p_m = B^2/2 \mu_0$, $\bm{F}$ is a  body force and
\begin{equation}
\tau_{ij} = \mu\left(\upartial_ i u_j + \upartial_ j u_i - \textstyle{\frac{2}{3}} \displaystyle \delta_{ij} \upartial_ k u_k \right),
\qquad \qquad
\Phi = \tau_{ij}\upartial_ i u_j.
\end{equation}
The specific heat at constant pressure $c_p$, the permeability $\mu_0$, the magnetic diffusivity $\eta$, the thermal conductivity $K$, the gas constant $R$ and the dynamic viscosity $\mu$ are all taken as constant. Although we shall assume a perfect gas throughout, the main ideas of the paper still hold for a more general equation of state.

An important point to make is that our analysis proceeds via three distinct stages. First we consider a purely hydrostatic, $z$-dependent \textit{reference state}. This is then perturbed by the inclusion of a horizontal magnetic field and aligned shear flow, both $z$-dependent, leading to a $z$-dependent MHD \textit{basic state}. Finally, we consider three-dimensional, time-dependent perturbations of this basic state. 

The reference state, consisting of a vertically stratified layer of gas in hydrostatic balance in the region $0 < z <d$, is governed by the equation
\begin{equation}
\frac{\drm \hp}{\drm z} = - g \hr,
\label{eq:pbalance}
\end{equation}
where $\hp$ and $\hr$ are the reference state pressure and density respectively. For any field variable $f$ we define the inverse scale height of a reference state $\hat{f}(z)$ by $H_f^{-1} = \drm ( \ln \hat{f}) / \drm z\big|_{z=0}$ and take $f_* = \hat{f}(0)$ to be a characteristic value of the variable. The physical idea of the Boussinesq approximation is that the depth of the layer $d$ is considered small in comparison with the pressure scale height, $H_p = c_s^2/g$, where the isothermal sound speed $c_s$ is defined by $c_s^2 = p_*/\rho_*$; note, from the equation of state~\eqref{eq:MHD_f}, that the density and temperature scale heights have the same magnitude as $H_p$ and so it follows that $d \ll H_\rho, H_T$. 

The reference state is modified by the introduction of a steady, horizontal magnetic field and an aligned steady shear flow. The field takes the form $\bm{B}_0 = B_0(z) \bfxhat$, where, for non-zero magnetic diffusivity, $B_0(z)$ is a linear function of height $z$; the flow $\bm{U}_0 = U_0(z) \bfxhat$ results from the (arbitrary) body force $\bfF$. These, in turn, introduce a perturbation of the reference state to form a \textit{basic state}. Analogous to the hats denoting reference state quantities, we shall use a subscript zero to denote the perturbations away from the reference state that result from the imposed magnetic field and shear flow. We define the scale heights $H_B$ and $H_U$ in terms of $B_0$ and $U_0$; at this stage we stipulate only that $d \lesssim H_B \lesssim H_p$ and $d \lesssim H_U \lesssim H_p$.  As the representative value for the magnetic field, we may take $B_*$ to be the rms value of $B_0(z)$ over the layer. For the velocity field, the physics is of course unchanged by the addition of a constant flow to $U_0(z)$; thus we define $U_*$ as the rms value of a shear flow in a frame of reference chosen such that the flow has zero mean. We make the assumption, as in \citet{SW_1982}, that the Alfv\'en speed $c_A = B_*/\sqrt{\mu_0 \rho_*}$ is small in comparison with the sound speed $c_s$; this guarantees that the difference between the reference and basic states is small. On subtracting off the reference state, the `$0$' variables satisfy the equations
\begin{subequations}
\begin{align}
\frac{\drm \varPi_0}{\drm z} &= -g\rho_0,\\
\frac{\drm^2 B_0}{\drm z^2} &= 0,\\
K\frac{\drm^2 T_0}{\drm z^2} &= - \frac{\eta}{\mu_0} \left( \frac{\drm B_0}{\drm z} \right)^2 - \mu \left( \frac{\drm U_0}{\drm z} \right)^2.
\end{align}
\end{subequations}

We now consider time-dependent, typically three-dimensional perturbations to the basic state. On denoting this perturbation of a field variable $f$ by $\delta \! f(\bm{x}, t)$, we may write
\begin{equation}
f(\bm{x},t) = \hat{f}(z) + f_0(z) + \delta \! f(\bm{x}, t),
\end{equation}
thus expressing $f$ in the terms of its reference state (hat), the steady perturbation arising from the imposed field and flow (subscript zero), leading to a basic state, and time-dependent perturbations away from the basic state (denoted by $\delta$).

On defining $\Delta f = f_0(d)-f_0(0)$ as the change in $f_0$ across the layer, we make the assumption \citep[cf.][]{SV_1960, Corfield_1984} that the size of the time-dependent perturbations does not exceed that of the jump across the layer, i.e.\ $\delta \! f = \mathrm{O}(\Delta f)$. Furthermore, for vector fields $\bff$, it is convenient to introduce the notation $f_{\parallel}$ and $f_{\perp}$, representing the magnitudes of the components of the fluctuations parallel and perpendicular  to the basic state magnetic field.

We proceed in a similar fashion to \cite{Corfield_1984} by finding appropriate magnitudes for the perturbations in terms of the basic state. With our focus on buoyancy-driven instabilities, an appropriate ordering is that the kinetic energy of the transverse flow results from buoyancy perturbations, i.e.
\begin{equation}
\rho_* \, \dvper^2 \sim \delta \! \rho \, g d.
\label{eq:energy}
\end{equation}
In the hydrodynamic Boussinesq approximation \citep{SV_1960}, fluctuations in gas pressure are small, the predominant balance in the equation of state being between temperature and density fluctuations.  The idea underlying magnetic buoyancy is that it is fluctuations in \textit{total} pressure that are considered small, with fluctuations in gas pressure therefore being comparable with those of magnetic pressure; thus gas pressure variations are retained in the perturbed equation of state. With this in mind, we adopt the same scaling for total pressure fluctuations as \citet{SW_1982}, namely
\begin{equation}
\delta \varPi \sim \delta \rho \, g d \sim \frac{\delta \rho}{\rho_*}\frac{d}{H_p}p_* ,
\label{eq:pi}
\end{equation}
from which it follows that
\begin{equation}
\frac{\delta p}{p_*} = - \frac{\delta p_m}{p_*} + \mathrm{O} \left( \frac{\delta \rho}{\rho_*}\frac{d}{H_p} \right).
\label{eq:pm}
\end{equation}
Thus the density perturbation may be expressed in terms of temperature and magnetic pressure perturbations as 
\begin{equation}
\frac{\delta \rho}{\rho_*} =  -\left(\frac{\delta T}{T_*} + \frac{\delta p_m}{p_*} \right) \left( 1 + \mathrm{O} \left( \frac{d}{H_p} \right) \right) .
\label{eq:state}
\end{equation}
On the assumption that the magnitude of the magnetic field fluctuations does not exceed that of the imposed field, the magnetic pressure perturbation may thus be written as
\begin{equation}
\delta p_m \approx \frac{B_* \, \Bpara}{\mu_0} \sim \frac{\delta \rho}{\rho_*}p_* .
\label{eq:mpressure}
\end{equation}
Balancing the two terms of the parallel component of $\bmu {\bm \cdot} \bmnabla \bmB$ provides the following crucial ordering:
\begin{equation}
\frac{\dvper \, \Bpara}{d} \sim \frac{\dvper \, B_*}{H_B}, \quad \mbox{implying} \quad \Bpara \sim \frac{d}{H_B} B_* .
\label{eq:mscale}
\end{equation}
Hence, using~\eqref{eq:mpressure} and \eqref{eq:mscale}, we obtain a relation between the magnitude of the density perturbations and that of the basic state magnetic field,
\begin{equation}
\frac{\delta \rho}{\rho_*}\frac{H_B}{d} p_* \sim \frac{B_*^2}{\mu_0} .
\label{eq:strB}
\end{equation}
Combining  the orderings (\ref{eq:strB}) and  (\ref{eq:energy}) then provides the consistent scaling of the magnitude of the perpendicular velocity in terms of the basic state magnetic field,
\begin{equation}
\frac{\dvper^2}{c_A^2}  \sim \frac{d}{H_B} \, \frac{d}{H_p}.
\label{eq:uperp_1}
\end{equation}
As shown by \citet{SW_1982} and \citet{Corfield_1984}, a significant difference between the standard Boussinesq equations and the magneto-Boussinesq equations is that the latter necessarily impose a restriction on the perturbation lengthscale $L$ in the direction of the imposed magnetic field. We now address this issue within our derivation; the arguments advanced to date are valid irrespective of the value of $L$.

We expect advection and stretching of the magnetic field to be of comparable importance; from the perpendicular and parallel components of the induction equation this gives the scalings
\begin{equation}
\frac{\dvper \, \Bper}{d} \sim \frac{B_* \, \dvper }{L} \quad \mbox{and} \quad \frac{\dvper \, \Bpara}{d} \sim \frac{B_* \, \dvpara }{L},
\label{eq:Bstar}
\end{equation}
leading, after the use of (\ref{eq:mscale}), to
\begin{equation}
\Bper \sim \frac{H_B}{L} \Bpara  \quad  \mbox{and} \quad \dvper \sim \frac{H_B}{L} \dvpara .
\label{eq:para}
\end{equation}

Finally, we use the fact that it is physically important to include the effects of magnetic tension. Balancing inertia against magnetic tension in the momentum equation leads to the ordering
\begin{equation}
\rho_* \frac{\dvper^2}{d} \sim \frac{B_*}{\mu_0} \frac{\Bper}{L},
\label{eq:uperp_2}
\end{equation}
and hence, using \eqref{eq:para}, to
\begin{equation}
\frac{\dvper^2}{c_A^2} \sim \frac{d^2}{L^2}.
\label{eq:uperp_3}
\end{equation}
In deriving \eqref{eq:uperp_2} we have used the perpendicular component of the momentum equation directly; balancing the terms in the parallel component and using the expressions~\eqref{eq:mscale} and \eqref{eq:para} for $\Bpara$ and $\dvpara$ respectively leads to the same result. Finally, combining the scalings~\eqref{eq:uperp_1} and \eqref{eq:uperp_3} provides an important constraint on the horizontal lengthscale, namely
\begin{equation}
L^2 \sim H_pH_B.
\label{eq:sizeL}
\end{equation}

The above scalings have been derived solely by consideration of the basic ideas of magnetic buoyancy, without any reference as yet to the shear flow $U_0(z)$. Their derivation follows a rather different line of argument to that of \citet{Corfield_1984}, in the process demonstrating their validity for magnetic field scale heights in the entire range $d \lesssim H_B \lesssim H_p$. In the case of $H_B \sim H_p$, they are entirely consistent with those of \citet{Corfield_1984}. As we shall see, for our future exposition involving the introduction of velocity shear, it is important that we make no \textit{a priori} assumption about the magnitude of $H_B$.

\section{Incorporating velocity shear \label{sec3}}

On demanding that velocity shear enters the momentum equation in a significant manner, a balance between inertia and magnetic tension gives
\begin{equation}
\rho_* \frac{\dvper \, U_*}{H_U} \sim \frac{1}{\mu_0}\frac{B_*\, \Bpara }{L} .
\label{eq:Ubalance1}
\end{equation}
Similarly, from  the induction equation, a balance between advection and stretching of magnetic field leads to
\begin{equation}
\frac{\dvper \, B_* }{H_B} \sim \frac{\Bper \, U_* }{H_U}.
\label{eq:Ubalance2}
\end{equation}
Equating these two expressions for $\dvper$, and making use of the orderings \eqref{eq:mscale} and \eqref{eq:para} for the relative sizes of the magnetic field perturbations, yields the important result,
\begin{equation}
\frac{U_*^2}{c_A^2} \sim \frac{H_U^2}{H_B^2}.
\label{eq:balanceUB}
\end{equation}

In Appendix~\ref{ap:noshear}, we consider the linear stability analysis of the basic state formed by the imposition of a shear flow into the magneto-Boussinesq equations of \citet{Corfield_1984}, i.e.\ with $L \sim H_B \sim H_p$; the flow is assumed to have scale height $H_U \sim d$ and a characteristic velocity comparable in magnitude with that of the velocity fluctuations. As such, $(\bm{U}_0 {\bm \cdot} \bmnabla)$ is neglected in favour of $(\bm{u}_\perp {\bm \cdot} \bmnabla)$ in the advective terms, although the shear (through $\bm{U}_0'$)  does appear in both the momentum and induction equations (\eqref{eq:A_momentum} and \eqref{eq:A_induction}). However, somewhat surprisingly, it plays no role in the resulting eigenvalue problem. Similarly, on adopting the scaling $H_U \sim L \sim H_B \sim H_p$, a Galilean transformation can be made such that the system in Appendix~\ref{ap:noshear} is again recovered. Thus the na\"ive introduction of a shear flow into the magneto-Boussinesq equations does not describe a regime in which the flow can influence the onset of magnetic buoyancy instabilities.

In order to involve the velocity shear in a meaningful manner, it is imperative therefore that two conditions are met. The first is that
\begin{equation}
H_U \sim d .
\label{eq:HUd}
\end{equation}
The second is that the imposed flow is significant in the advective terms; this requires that $(\bm{U}_0 {\bm \cdot} \bmnabla)$ and $(\bm{u}_\perp {\bm \cdot} \bmnabla)$ be of comparable magnitude, thus forcing a balance between the basic state velocity and the perpendicular velocity perturbation,
\begin{equation}
\frac{U_*}{L} \sim \frac{\dvper}{d}.
\label{eq:velos}
\end{equation}
Henceforth, we shall refer to a shear flow that satisfies both \eqref{eq:HUd} and \eqref{eq:velos} as being \textit{influential}. Combining the two expressions for $\dvper$, \eqref{eq:uperp_3} and \eqref{eq:velos}, provides the following important ordering for the magnitude of the shear flow in terms of the Alfv\'en velocity of the imposed magnetic field,
\begin{equation}
U_*^2 \sim c_A^2.
\label{eq:U_cA}
\end{equation}

Whereas the scalings of section~\ref{sec2} are valid for magnetic field scale heights satisfying $d \lesssim H_B \lesssim H_p$, the requirement that the imposed shear flow influences the magnetic buoyancy instability places a tighter restriction on $H_B$. Scalings~\eqref{eq:balanceUB}, \eqref{eq:HUd} and \eqref{eq:U_cA} lead to the crucial result that
\begin{equation}
H_B \sim H_U \sim d.
\label{eq:HBd}
\end{equation}

\section{\label{sec4} The magneto-Boussinesq velocity shear equations}

\subsection{Derivation of the equations}\label{sec4_1}

Sections~\ref{sec2} and \ref{sec3} provide the framework required to introduce velocity shear into the magneto-Boussinesq approximation. We shall now incorporate these ideas into the derivation of an asymptotically consistent set of governing equations. We focus on an influential shear flow, with $H_U \sim H_B \sim d$, and define two small parameters,
\begin{equation}
\ep = \frac{d}{H_p} \quad  \mbox{and} \quad  \ept = \frac{c_A^2}{c_s^2},
\label{eq:eps}
\end{equation}
where $\ep,\ept \ll  1$. Physically, $\ep$ is a measure of the inverse pressure scale height of the hydrostatic reference state, whereas $\ept$, through \eqref{eq:strB}, provides a measure of the amplitude of the fluctuations driven by magnetic buoyancy. (We note that our $\ept$ is of the same order of magnitude as the $\ept$ of \cite{Corfield_1984}, defined as $\delta \rho /\rho$.) It follows from \eqref{eq:U_cA} that
\begin{equation}
U_*^2 \sim \ept c_s^2.
\label{eq:Ucs}
\end{equation}
Using expression~\eqref{eq:sizeL}, we may rewrite the horizontal lengthscale in terms of $\ep$,
\begin{equation}
\frac{d}{L} \sim {\ep}^{1/2}.
\label{eq:dL}
\end{equation}

We non-dimensionalise $T$ by $T_*$, $p$ by $p_*$, $\rho$ by $\rho_*$, $p_m$ by $p_*$, lengths with $d$ and times with the sound crossing time across the layer. The condition that motion-induced fluctuations do not exceed, in order of magnitude, static variations across the layer translates to the requirement that $\ept \lesssim \ep$. Following \citet{Mal_1964} and \citet{Corfield_1984}, we express all variables in terms of the two small parameters, with non-dimensional variables of order unity denoted by a tilde. Based on the scalings derived in section~\ref{sec2}, the thermodynamic quantities are expressed as
\begin{subequations}
\label{eq:expan}
\begin{align}
T \left (\mathbf{x} , t \right) &=  T_* \left( 1 + \ep \frac{H_p}{H_T} \frac{z}{d} + \ept \overset{\sim}{T}_0 + \ept \delta \overset{\sim}{T} \left( \mathbf{x}, t \right) + \dots \right) , \label{eq:expan_a}\\
p \left( \mathbf{x}, t \right) &= p_* \left( 1 + \ep\frac{z}{d}  +\ept\overset{\sim}{p}_0 + \ept \delta \overset{\sim}{p} \left( \mathbf{x}, t \right)  + \dots \right) ,\\
\rho \left( \mathbf{x}, t \right) &= \rho_* \left( 1 + \ep \frac{H_p}{H_\rho}\frac{z}{d} + \ept \overset{\sim}{\rho}_0 + \ept \delta \overset{\sim}{\rho} \left( \mathbf{x},t \right) + \dots \right) ,
\end{align}
where we have linearised the reference state. The magnetic and total pressure are expanded as
\begin{align}
p_m(\mathbf{x},t) &=\ept p_*(\overset{\sim}{p}_{m0} + \delta \overset{\sim}{p}_m(\mathbf{x},t)  + \dots),\\
\varPi(\mathbf{x},t) &= \ept p_*(\overset{\sim}{\varPi}_{0} + \ep \delta \overset{\sim}{\varPi}(\mathbf{x},t)  + \dots), \label{eq:expan_e}
\end{align}
where expression~\eqref{eq:pi} has been used for the ordering of the $\delta \overset{\sim}{\varPi}$ term.

It is convenient to split the velocity and magnetic field into their parallel and perpendicular components; from expressions~\eqref{eq:para}, \eqref{eq:Ucs} and \eqref{eq:dL} these become
\begin{align}
\bm{u} &=  \ept^{1/2} c_s \left(\overset{\sim}{\bm{U}}_0 + \delta \overset{\sim}{\bm{u}}_\parallel+\ep^{1/2}\delta \overset{\sim}{\bm{u}}_\perp\right),\\
\bm{B} &= (\ept \mu_0 p_*)^{1/2} \left(\overset{\sim}{\db} + \delta\overset{\sim}{\bm{B}}_\parallel + \ep^{1/2} \delta \overset{\sim}{\bm{B}}_\perp\right). \label{eq:B_scaling}
\end{align}
Based on \eqref{eq:dL}, we write
\begin{equation}
\bmnabla_\parallel = \frac{\ep^{1/2}}{d} \overset{\sim}{\bmnabla}_{\parallel}, \qquad  \bmnabla_\perp     = \frac{1}{d}               \overset{\sim}{\bmnabla}_{\perp}.
\end{equation}
The time scale is determined by the conventional Boussinesq approach of balancing the vertical acceleration against the buoyancy. Using the scalings for $\delta \rho$ and $\dvper$, this gives
\begin{equation}
\upartial_t = \left( \ep\ept \right)^{1/2} \frac{c_s}{d} \upartial_{\overset{\sim}{t}} .
\label{eq:time}
\end{equation}
\end{subequations}

The various expansions~\eqref{eq:expan} are then substituted into the MHD equations~\eqref{eq:MHD}. To simplify the notation, we drop the tildes, write $\delta \bfB = \bfb$ and drop the $\delta$ from the other terms. After substituting for $\delta p$ and $\delta \rho$ from equations~\eqref{eq:pm} and \eqref{eq:state}, and removing terms that arise from the basic state, the governing equations at leading order become:
\begin{subequations}
\label{eq:new}
\begin{equation}
\bmnabla {\bm \cdot} \bm{u} = 0, \label{eq:new_a}
\end{equation}
\begin{equation}
{\bm \nabla} {\bm \cdot} \bm{b} = 0, \label{eq:new_b}
\end{equation}
\begin{align}
\nonumber \left( \upartial_t \, +\, (\bm{U}_0 + \bm{u}) {\bm \cdot} \bmnabla \right) \bm{u} + w \upartial_z \bm{U}_0 = &- \per \varPi \, +\, \left( T + p_m\right) \bfzhat \\
&+ \,  \left(\db + \bm{b} \right) {\bm \cdot} \bmnabla \bm{b} +  b_z \upartial_z \db \,+\, (\sigma/R_a)^{1/2}\nabla_\perp^2 \bm{u},
\label{eq:new_c}
\end{align}
\begin{align}
\nonumber \left( \upartial_t \, +\, (\bm{U}_0 +\bm{u}) {\bm \cdot} \bmnabla \right) \bm{b} + w \upartial_z \db= \,   &\left( \db + \bm{b} \right) {\bm \cdot} \bmnabla \bm{u}  \\
&+ b_z \upartial_z \bm{U}_0  +\,  \sigma_m^{-1} \left( \sigma/ R_a \right)^{1/2} \nabla_\perp^2 \bm{b},
\label{eq:new_d}
\end{align}
\begin{align}
\nonumber \left( \upartial_t \, +\, (\bm{U}_0 +\bm{u}) {\bm \cdot} \bmnabla \right) \left( T_0 + T \right) &+ D \left( \upartial_t \, +\, (\bm{U}_0 +\bm{u}) {\bm \cdot} \bmnabla \right) \left( -p_0 + p_m \right) +   w\beta = (\sigma R_a)^{-1/2} \nabla_\perp^{2} T \\
\nonumber &+ D \sigma_m^{-1} \left( \sigma/ R_a \right)^{1/2}  \left( (\upartial_y b_\parallel)^2 + (\upartial_z b_\parallel )^2 + 2 \upartial_z B_0 \upartial_z b_\parallel \right) \\
&+ D (\sigma/R_a)^{1/2} \left( (\upartial_y u_\parallel)^2 + (\upartial_z u_\parallel )^2 + 2 \upartial_z U_0 \upartial_z u_\parallel \right),
\label{eq:new_e} 
\end{align}
\end{subequations}
where the vertical components of the velocity and magnetic field perturbations are denoted by $w$ and $b_z$ respectively. It is worth noting that from the scaling~\eqref{eq:B_scaling}, $p_m$ in equations~\eqref{eq:new} is given by $p_m = B_0 b_\parallel + b_\parallel^2/2$; the perturbation to the total pressure is $\varPi = p + p_m$. The operator $\bmnabla$ is defined as
\begin{equation}
\bmnabla = \bmnabla_\parallel + \bmnabla_\perp.
\end{equation}
 The various non-dimensional numbers are defined as follows:
\begin{equation}
\sigma = \frac{\nu}{\kappa}, \quad \sigma_m = \frac{\nu}{\eta}, \quad Ra = \ept \frac{g d^3}{\nu \kappa} , \quad D = \frac{\gamma - 1}{\gamma} ,
\end{equation}
where $\gamma$ is the conventional ratio of specific heats; $\sigma$ is the Prandtl number, $\sigma_m$ is the magnetic Prandtl number and $Ra$ is the Rayleigh number (note that our $D$ is equivalent to $D^{-1}$ in \cite{Corfield_1984}). Ensuring that the diffusion terms do not dominate imposes the restriction that $(\sigma/R_a)^{1/2}$, $\sigma_m^{-1} \left( \sigma/ R_a \right)^{1/2}$ and $(\sigma R_a)^{-1/2}$ are all $\Orm (1)$. For asymptotic consistency, the subadiabatic temperature gradient in equation~\eqref{eq:new_e} must be $\mathrm{O} (\ept)$, and so we have defined
\begin{equation}
\frac{1}{\gamma} \frac{\drm}{\dz} \ln \left( \frac{\hp}{\hr^\gamma} \right) = \frac{\ept \beta}{d}.
\end{equation}

Equations~\eqref{eq:new} are derived only under the assumption that $\ept \lesssim \ep$. If $\ep$ and $\ept$ are comparable then the subadiabatic gradient is $\mathrm{O} (\ep)$, comparable in magnitude with its component gradients of pressure and density. However, if $\ept \ll \ep$ then the subadiabatic gradient, being $\mathrm{O} (\ept)$, is formally smaller than the pressure and density gradients, and therefore in this case, equations~\eqref{eq:new} hold only for atmospheres that are close to adiabatic. Finally we note that equations~\eqref{eq:new} may be expressed in an alternative form through the introduction of the variable $\bm{V} = \bm{U}_0 + \bm{u}$; this leads to a certain simplification, through the combination of terms, though the dissipation of the basic state velocity $\bm{U}_0$ must then be accounted for in equations~\eqref{eq:new_c} and \eqref{eq:new_e}.

\subsection{Comparison with the equations of \citet{SW_1982} and \citet{Corfield_1984}} \label{sec4_2}

There are significant differences between our new system of equations~\eqref{eq:new} and the equations derived by \citet{SW_1982} and \citet{Corfield_1984}. We have shown that in order to maintain consistent scalings following the introduction of an influential shear flow, the magnetic field scale height $H_B$ has to be $\mathrm{O} (d)$, considerably smaller than that adopted in \cite{Corfield_1984}, namely $H_B \sim H_p$. Through the scalings derived in sections~\ref{sec2} and \ref{sec3}, this leads to important differences in the magnitudes of both perturbation and basic state quantities.

The \citet{Corfield_1984} ordering of $H_B \sim H_p$ forces $L \sim H_p$ through expression~\eqref{eq:sizeL}; in turn, from \eqref{eq:para}, this implies that for both the flow and field perturbations, the perpendicular and parallel components have the same magnitude. This is in marked contrast to our system, in which although the perpendicular components of the flow and field scale as in \citet{Corfield_1984}, namely
\begin{equation}
\dvper^2 \sim \ep \ept c_s^2 \qquad \mathrm{and} \qquad \frac{\Bper^2}{\mu_0} \sim \ep \ept p_* ,
\end{equation}
the parallel components are $\mathrm{O}(\ep^{-1/2})$ greater. Furthermore, from \eqref{eq:Bstar}, the characteristic strength of the basic state magnetic field $B_*$ is given by
\begin{equation}
\frac{B_*^2}{\mu_0} \sim \frac{L^2}{d^2} \frac{\Bper^2}{\mu_0} \sim \frac{L^2}{d^2} \ep \ept p_* ,
\label{eq:Bstar_2}
\end{equation}
thus highlighting a further important difference between the two systems: for our equations, governed by the scaling $L^2 \sim d H_p$,  expression \eqref{eq:Bstar_2} becomes
\begin{equation}
\frac{B_*^2}{\mu_0} \sim \ept p_* ,
\end{equation}
whereas for \citet{Corfield_1984} the characteristic field strength $B_*$ is $\mathrm{O}\left( \ep^{-1/2} \right)$ greater. Hence the condition that an imposed shear flow be influential requires an $\mathrm{O} \left( \ep^{1/2} \right)$ reduction in the strength of the basic state magnetic field.

Unlike the equations of \citet{SW_1982}, equations~\eqref{eq:new} now satisfy, at leading order, both the full incompressibility condition~\eqref{eq:new_a} and the full solenoidal condition on the magnetic field~\eqref{eq:new_b}. Consequently, since the new system is fully incompressible, there is no longer a next-order correction of $\bm{\nabla} \, {\bm \cdot} \, \bm{u}$ to the induction equation and hence the induction equation now conserves $\bm{\nabla}\, {\bm \cdot}\, \bm{b}$. Note also that, in contrast to the standard Boussinesq approximation, both Ohmic and viscous heating terms are included in the energy equation~\eqref{eq:new_e}, these terms arising as a consequence of having increased the magnitude of both the parallel velocity and parallel magnetic field perturbations.

\section{Linking the magneto-Boussinesq systems \label{sec5}}
In the previous section, we derived a new set of equations describing the combined effects of magnetic buoyancy instability and an influential shear flow, consistent within the magneto-Boussinesq approximation. As noted above, there are a number of significant differences between these equations and those of \cite{SW_1982}. It is therefore of interest to examine how a connection may be made between the two systems. In order to do this, we again fix $H_U \sim d$, but choose not to impose the conditions of an influential shear, \eqref{eq:HUd} and \eqref{eq:velos}. This then allows us to introduce a control parameter $q$, defined by
\begin{equation}
\frac{H_B}{H_p} = \ep^q,
\end{equation}
where $q$ satisfies $0\le q \le 1$ and is a measure of the relative sizes of the scale heights of magnetic field and pressure. Varying $q$ then allows us to examine how the system of equations changes from when $q=0$ \citep{SW_1982, Corfield_1984}, in which the velocity shear has no effect on the onset of instability (see Appendix~\ref{ap:noshear}), to when $q=1$, the case considered in section~\ref{sec4}. In order to keep the magnitudes of the density fluctuations constant as the parameter $q$ varies, we consider, using \eqref{eq:strB}, basic state magnetic fields of strength
\begin{equation}
\frac{B_*^2}{\mu_0 p_*} = \ep^{q-1} \ept .
\label{eq:scaling2}
\end{equation}
The assumption that the Alfv\'{e}n speed is much smaller than the sound speed leads to the inequality $\ept \ll \ep^{1-q}$. Following the ideas of sections~\ref{sec2} and \ref{sec3}, we can express the required strength of the velocity shear and horizontal length scale in terms of the parameter $q$ as
\begin{equation}
U_*^2 \sim \ep^{1-q} \ept c_s^2, \quad \frac{d}{L} \sim \ep^{1- q/2}.
\end{equation} 
Although more complicated than the $q=1$ expressions of section~\ref{sec4}, we can nonetheless proceed in a similar manner and express the variables in terms of non-dimensional expansions. The scalar variables are independent of $q$ and are therefore scaled as in \eqref{eq:expan_a} -- \eqref{eq:expan_e}; the vector quantities may be expanded as
\begin{subequations}
\begin{align}
\bm{u} &=  (\ep^{1-q}\ept)^{1/2} c_s \left(\overset{\sim}{\bm{U}}_0 + \delta \overset{\sim}{\bm{u}}_\parallel+\ep^{q/2}\delta \overset{\sim}{\bm{u}}_\perp\right),\\
\bm{B} &= (\ep^{q-1}\ept \mu_0 p_*)^{1/2} \left(\overset{\sim}{\db} + \ep^{1-q}\delta\overset{\sim}{\bm{b}}_\parallel + \ep^{1-q/2}\delta\overset{\sim}{\bm{b}}_\perp\right).
\end{align}
The operators $\bmnabla_\parallel$ and $\bmnabla_\perp$ are scaled as
\begin{equation}
\bmnabla_\parallel = \frac{\ep^{1-q/2}}{d} \overset{\sim}\bmnabla_\parallel, \qquad  \bmnabla_\perp     = \frac{1}{d}               \overset{\sim}\bmnabla_\perp ,
\end{equation}
and we adopt the same $q$-independent time scale as in \eqref{eq:time},
\begin{equation}
\upartial_t = \left( \ep\ept \right)^{1/2} \frac{c_s}{d} \upartial_{\overset{\sim}{t}} .
\end{equation}
\end{subequations}
Performing the same sequence of operations that lead to equations~\eqref{eq:new}, leads to the following $q$-dependent mixed-order system of equations:
\begin{subequations}
\label{eq:full}
\begin{equation}
\label{eq:full1}  - \ep \frac{w}{H_\rho} + \ep^{1-q} \para {\bm \cdot} \bm{u} + \per {\bm \cdot} \bm{u} = 0,
\end{equation}
\begin{equation}
\label{eq:full2} \ep^{1-q} \para {\bm \cdot} \bm{b} + \per {\bm \cdot} \bm{b} = 0,
\end{equation}
\begin{align}
\nonumber \left(\upartial_t + \ep^{1-q} (\bm{U}_0 + \bm{u}) {\bm \cdot} \para + \bm{u}{\bm \cdot} \per \right) \bm{u} + w \upartial_z \bm{U}_0 &= - \per \varPi + \left(  T+ p_m \right) \bfzhat \\
\label{eq:full3} + \left( \bm{B}_0 {\bm \cdot} \para + \ep^{1-q} \bm{b} {\bm \cdot} \para + \bm{b} {\bm \cdot} \per \right) \bm{b} &+ b_z \upartial_z \bm{B}_0 + ( \sigma/R_a)^{1/2} \nabla_{\perp}^2 \bm{u},
\end{align}
\begin{align}
\nonumber \left( \upartial_t + \ep^{1-q} (\bm{U}_0 + \bm{u}) {\bm \cdot} \para + \bm{u} {\bm \cdot} \per \right) \bm{b} &+ w \upartial_z \bm{B}_0 = \left( \bm{B}_0 {\bm \cdot} \para + \ep^{1-q} \bm{b} {\bm \cdot} \para + \bm{b} {\bm \cdot} \per\right) \bm{u} \\
\label{eq:full4}  &+ b_z \upartial_z \bm{U}_0 - \bm{B}_0 \left( \bmnabla {\bm \cdot} \bm{u} \right) + \sigma_m^{-1} \left( \sigma/ R_a \right)^{1/2} \nabper^2 \bm{b},
\end{align}
\begin{align}
\nonumber \left( \upartial_t + \ep^{1-q} (\bm{U}_0 \right. & \left. + \bm{u}) {\bm \cdot} \para + \bm{u} {\bm \cdot} \per \right) \left((T_0 + T) +  D \left( -p_0 + p_m \right) \right) + w \beta = \, (\sigma R_a)^{-1/2} \nabper^2 T \\
\nonumber &+  \ep^{1-q} D \sigma_m^{-1} \left( \sigma/ R_a \right)^{1/2} \left( (\upartial_y b_\parallel)^2+(\upartial_z b_\parallel)^2 +2\upartial_z B_0\upartial_z b_\parallel \right) \\
\label{eq:full5} &+ \ep^{1-q} D (\sigma/R_a)^{1/2} \left( (\upartial_y u_\parallel)^2+ ( \upartial_z u_\parallel )^2 + 2\upartial_z U_0\upartial_z u_\parallel \right). 
\end{align}
\end{subequations}

Note that special attention is needed when considering the $\bm{B}_0 ( \bmnabla {\bm \cdot} \bm{u} )$ term in \eqref{eq:full4}. In more detail, this term takes the form
\begin{equation}
\bm{B}_0 \left( \bmnabla_\parallel {\bm \cdot} \bm{u} + \ep^{q-1} \bmnabla_\perp {\bm \cdot} \bm{u} \right) ,
\end{equation}
which, on using \eqref{eq:full1}, can be written as
\begin{equation}
\ep^q  \bm{B}_0 \frac{w}{H_\rho} .
\label{eq:divu}
\end{equation}

When $q=0$, the term \eqref{eq:divu} enters equation~\eqref{eq:full4} at leading order, as in \citet{Corfield_1984}. This substitution can be performed, however, only for $q=0$; for all other values of $q$, this term is formally smaller than those involving $\para {\bm \cdot} \bm{u}$ and $\per {\bm \cdot} \bm{u}$. 

From equations~\eqref{eq:full}, three different systems can be identified, depending on the choice of $q$. For $q=0$ the system reverts to the magneto-Boussinesq equations of \citet{SW_1982}, for which the inclusion of a shear flow has no influence on the onset of instability (Appendix~\ref{ap:noshear}). The range $0<q<1$ produces a very similar system, but with no density term in the induction equation. For this system, via the same analysis as in Appendix~\ref{ap:noshear}, it can again be shown that the shear has no effect on the linearised diffusionless system of equations. The final system comes from taking $q=1$, thereby recovering equations~\eqref{eq:new}. It is important to remember that increasing $q$ essentially decreases the magnetic field scale height from $H_B \sim H_p$, where the variations across the layer are small in comparison with the uniform component of $B_0$, to $H_B \sim d$, where both the uniform component and the variations across the layer are of the same order.

\section{Discussion\label{sec6}}

The principal result of this paper is the derivation of a new set of MHD equations governing the evolution of magnetic buoyancy instabilities in the magneto-Boussinesq approximation and in the presence of a horizontal, depth-dependent shear flow. The equations are derived via an expansion procedure in two small parameters: $\ep$, the ratio of the layer depth to the pressure scale height, and $\ept$, the ratio of the square of the Alfv\'en speed to the square of the sound speed. Section~\ref{sec2}, which follows the treatment of \citet{Corfield_1984} to a certain extent, lays the foundations for the magneto-Boussinesq orderings in general, without specific reference to the incorporation of any shear flow. Unlike \citet{Corfield_1984} however, we make no assumption about the magnetic field scale height; as a result, all the orderings are valid for $d \lesssim H_B \lesssim H_p$. Section~\ref{sec3}, with reference to Appendix~\ref{ap:noshear},  describes how the na\"{i}ve incorporation of a shear flow into the equations of \citet{SW_1982} and \citet{Corfield_1984} (i.e.\ with $H_B \sim H_p$) has no influence on the linear stability problem. In order that the shear flow assumes a non-trivial role in the magnetic buoyancy instability, two conditions must be met: that $H_U$ is $\mathrm{O}(d)$ and that $U_*^2 \sim c_A^2$. For consistency with the scalings determined in section~\ref{sec2}, it follows that $H_B$ must also be $\Orm(d)$. The various orderings derived in sections~\ref{sec2} and \ref{sec3} are applied to the full MHD governing equations in section~\ref{sec4}, yielding the leading order equations~\eqref{eq:new}. Interestingly, equations~\eqref{eq:new} also allow us to examine the effects of magnetic buoyancy for a magnetic field with an $\mathrm{O} (d)$ scale height in the absence of velocity shear, a scenario that is excluded from the equations of \citet{Corfield_1984}, as identified by \citet{Hughes_1985}. The transformation between equations~\eqref{eq:new} and those of \citet{SW_1982} and \citet{Corfield_1984} can be effected by increasing the magnetic field scale height from $\mathrm{O}(d)$ to $\mathrm{O}(H_p)$. Section~\ref{sec5} describes the resulting changes in the governing equations, identifying three different regimes, each with their own set of equations: $H_B \sim d$ (equations~\eqref{eq:new}), $H_B \sim H_p$ (\citet{SW_1982} and \citet{Corfield_1984}) and a third, intermediate regime.

Finally, it is important to consider the implications of our study to the solar tachocline, and, in particular, to examine the parameter regimes in which the set of equations~\eqref{eq:new} is expected to hold. Let us first consider the magnitudes of the two small quantities in our asymptotic expansions, $\ep$ and $\ept$. The pressure scale height in the tachocline $\approx 0.08 R_\odot$ \citep{Gough_2007}. Estimates of the vertical extent of the tachocline vary a little, according to how it is defined \citep[see, for example,][]{Miesch_2005}, but lie in the range between $0.02 R_\odot$ and $0.05 R_\odot$. Thus $\ep = d/H_p$ is certainly less than unity, but is not particularly small. As for the ratio $\ept$, this is $\mathrm{O} \left( 10^3 / B_*^2 \right)$, where $B_*$ is measured in Gauss \citep{Ossendrijver_2003}. Estimates of the mean toroidal field strength in the tachocline result solely from theoretical considerations, and vary between $10^3 \Grm$ and $10^5 \Grm$, depending on the theoretical assumptions involved; this certainly makes $\ept$ small, in the range $10^{-7} \lesssim \ept \lesssim 10^{-3}$.

Given that the magnitudes of $\ep$ and $\ept$ for the tachocline suggest, at least \textit{a priori}, that a magneto-Boussinesq approach is appropriate, we nonetheless need to examine whether the tachocline shear flow inferred from helioseismology is \textit{influential} in the sense of equations~\eqref{eq:HUd} and \eqref{eq:U_cA}. Equation~\eqref{eq:HUd} specifies that $H_U \sim d$; this is true of the tachocline, almost by definition. Expression~\eqref{eq:U_cA} requires that $U_*$ be comparable with the Alfv\'en speed $c_A$. Since we have a good estimate of $U_*$ from helioseismological inversions, but no direct knowledge of the magnetic field strength $B_*$, it makes more sense to look at this from the other perspective and to ask what values of $B_*$ will allow \eqref{eq:U_cA} to be satisfied. From the helioseismological results of \citet{Schou_etal_1998}, the jump in the angular velocity across the tachocline (at the equator) is of the order of 20 nHz, which translates into $U_* \approx 30 \mathrm{ms}^{-1}$. Requiring $c_A \sim U_*$ determines the characteristic magnetic field strength as $B_* \approx 10^3 \Grm$. Thus everything ties together very nicely, suggesting that equations~\eqref{eq:new} form an appropriate system for the study of magnetic buoyancy instabilities in the tachocline.

\medskip

\noindent \textbf{Acknowledgements}

\smallskip

\noindent JAB was supported by an STFC studentship; DWH and EK were supported by the STFC Consolidated Grant ST/K000853/1.

\appendices

\section{\label{ap:noshear}}
The aim of this appendix is to demonstrate how introducing velocity shear in the `obvious' manner into the magneto-Boussinesq equations of \citet{SW_1982} and \citet{Corfield_1984} has no effect on the linear stability of the diffusionless system.

Suppose that we consider a basic state magnetic field of the form
\begin{equation}
\bm{B}_0 = B_* \left(1 - \frac{\lambda z}{d} \right) \hat{\bm{x}},
\end{equation}
where $\lambda = \mathrm{O} (d/H_p)$. In addition, we consider an aligned basic state velocity shear $\bm{U}_0 = U_0(z) \bfxhat$, with scale height $H_U$. We consider separately the two cases of $H_U = \mathrm{O} (d)$ and $H_U = \mathrm{O} (H_p)$.

\medskip

\noindent(i)\ $H_U \sim d$

\smallskip

We suppose that the flow $U_0(z)$ is an arbitrary function of $z$. As a consequence of the \citet{Corfield_1984} ordering of $L \sim H_B \sim H_p$, the advective terms are $\mathrm{O} (d/H_p)$ smaller than the shear terms and hence are neglected. On following \citet{SW_1982}, by linearising the governing equations, ignoring all diffusivities, and adopting $d$ as the unit of length and the Alfv\'{e}n period $d/c_A$ as the unit of time, we obtain the dimensionless equations,
\begin{align}
\upartial_t \bm{u} + w\bm{U}_0' &= -\per \varPi + \ep b_x \hat{\bm{z}} + \upartial_x \bm{b} - \lambda b_z \hat{\bm{x}},
\label{eq:A_momentum} \\
\upartial_t \bm{b} - \lambda w \hat{\bm{x}} &= \upartial_x \bm{u} + b_z \bm{U}_{0}' - \ep w \hat{\bm{x}},
\label{eq:A_induction}
\end{align}
where $\ep = d/H_p$, $\upartial_x = \mathrm{O} (\ep)$, and where, for simplicity, we have taken $\beta = 0$ in the energy equation and made the substitution $T = -p_m/(c_p\rho_*) $ (cf.\ equations~(40) and (39) in \citet{SW_1982}).  Since $\per {\bm \cdot} \bfu = 0$ and $\per {\bm \cdot} \bfb = 0$, we may introduce stream and flux functions, $\psi$ and $\chi$, such that
\begin{equation}
\bm{u} = \left(u, -\frac{\upartial \psi}{\upartial z}, \frac{\upartial \psi}{\upartial y} \right), \quad \bm{b} = \left(b_x, -\frac{\upartial \chi}{\upartial z}, \frac{\upartial \chi}{\upartial y} \right),
\end{equation}
where
\begin{equation}
u(x,y,z,t) = {\hat u}(z) \exp \left( \irm kx + \irm ly + pt \right)  , \ \textrm{etc.} \label{eq:uhat}
\end{equation}
Substituting expressions~\eqref{eq:uhat} into the $x$-component of the momentum equation \eqref{eq:A_momentum} and its curl yields, after dropping the hats,
\begin{align}
\label{eq:mo1} p u + \irm l U_0' \psi &= \irm k b_x - \irm l \lambda \chi , \\
\label{eq:mo2} p \left( -l^2 \psi + \psi'' \right)  &= \irm l \ep b_x + \irm k \left( -\l^2 \chi + \chi'' \right).
\end{align}
In a similar manner, the induction equation~\eqref{eq:A_induction} gives
\begin{align}
\label{eq:in1} p b_x - \irm l \lambda \psi &= \irm k u + \irm l \chi U_0'  - \irm l \ep \psi , \\
\label{eq:in2} \irm p \chi &=  -k \psi.
\end{align}
On eliminating $u$ between equations~\eqref{eq:mo1} and \eqref{eq:in1}, we obtain
\begin{equation}
\left( p^2 + k^2 \right) b_x = k l U_0' \psi + \irm l ( \lambda - \ep) p \psi + k l \lambda \chi + \irm l p U_0' \chi ,
\end{equation}
which, after substituting for $\chi$ from \eqref{eq:in2}, becomes
\begin{equation}
p \left( p^2 + k^2 \right) b_x = \irm l ( \lambda - \ep) p^2 \psi + \irm k^2 l \lambda \psi.
\label{eq:A_predisp}
\end{equation}
Equation~\eqref{eq:mo2}, after substituting for $\chi$ from \eqref{eq:in2}, and equation~\eqref{eq:A_predisp} form an eigenvalue problem for $p$ involving only the functions $\psi$ and $b_x$. The crucial point to note is that $U_0'$ does not appear in these expressions; hence the shear has no influence on the growth rate $p$.

\medskip

\noindent(ii)\ $H_U \sim H_p$

\smallskip

If the scale height $H_U$ is comparable with $H_p$ and $H_B$ then the major change to equations~\eqref{eq:A_momentum} and \eqref{eq:A_induction} is that the advective terms come into play. However, since $U_0(z)$ now varies on a scale very large compared with $d$, this simply represents, to a first approximation, advection by a uniform flow. A straightforward transformation therefore recovers equations~\eqref{eq:A_momentum} and \eqref{eq:A_induction} (with $U_0'$ now treated as a constant). Thus, with the ordering $H_U \sim H_p \sim H_B$, the imposed shear once again has no bearing on the linear stability problem.

\end{document}